\documentclass[epj]{webofc}
\usepackage[utf8]{inputenc}
\usepackage[varg]{txfonts}   % Web of Conferences font
\usepackage{booktabs}
\usepackage{xcolor}
\definecolor{darkred}{rgb}{0.4,0.0,0.0}
\definecolor{darkgreen}{rgb}{0.0,0.4,0.0}
\definecolor{darkblue}{rgb}{0.0,0.0,0.4}
\usepackage[bookmarks,linktocpage,colorlinks,
    linkcolor = darkred,
    urlcolor  = darkblue,
    citecolor = darkgreen]{hyperref}
\usepackage{subfigure}
\wocname{EPJ Web of Conferences}
\woctitle{Lattice2017}
\begin{document}
%%%%%%%%%%%%%%%%%%%%%%%%%%%%%%%%%%%%%%%%%%%%%%%%%%%%%%%%%%%%%%%%%%%%%%%%%%%%%
%
\selectlanguage{english}
%----------------------------------------------------------------------------
\title{%
Parton distribution functions on the lattice and in the continuum
}
%----------------------------------------------------------------------------
\author{%
\firstname{Joseph} \lastname{Karpie}\inst{1,2}%\fnsep\thanks{Acknowledges support from} 
\and
\firstname{Kostas} \lastname{Orginos}\inst{1,2}\fnsep\thanks{Speaker, \email{kostas@wm.edu}} \and
\firstname{Anatoly}  \lastname{Radyushkin}\inst{2,3}\and
\firstname{Savvas}  \lastname{Zafeiropoulos}\inst{1,2,4}
% etc.
}
%----------------------------------------------------------------------------
\institute{%
Department of Physics, The College of William \& Mary, Williamsburg, VA 23187, USA
\and
Thomas Jefferson National Accelerator Facility, Newport News, VA 23606, USA
\and
Physics Department, Old Dominion University, Norfolk, VA 23529, USA
\and
Institute for Theoretical Physics, Universit\"at Heidelberg, Philosophenweg 12, D-69120 Germany
}
%----------------------------------------------------------------------------
\abstract{%
 Ioffe-time distributions, which are functions of the Ioffe-time $\nu$,  are the Fourier transforms of parton distribution functions with respect to the  momentum fraction variable $x$.
 These distributions can be obtained from suitable equal time, quark bilinear hadronic  matrix elements which can be calculated from first principles in lattice QCD, as it has been recently argued. In this talk I present the first numerical calculation of the Ioffe-time distributions of the nucleon in the quenched approximation.
}
%----------------------------------------------------------------------------
\maketitle
%----------------------------------------------------------------------------
\section{Introduction}\label{intro}

The introduction of quasi parton distribution functions (quasi-PDFs) by X. Ji~\cite{Ji:2013dva} opened a new avenue for non-perturbative calculations of parton distribution functions (PDFs) circumventing a number of difficulties  Euclidean lattice QCD calculations had to face. As discussed in detail in~\cite{Ma:2014jla,Ma:2014jga,Ma:2017pxb}, one computes in lattice QCD certain hadronic matrix elements
that are related to parton distribution functions (PDFs) from which  PDFs can be extracted in a similar manner with which PDFs are extracted from experimental crossections.
 Several works  numerically implementing  these ideas~\cite{Lin:2014zya,Chen:2016utp,Alexandrou:2015rja,Zhang:2017bzy,Alexandrou:2017qpu} have already appeared in the literature. Methods for renormalizing these matrix elements have been constructed~\cite{Monahan:2016bvm,Alexandrou:2017huk, Chen:2017mzz, Ishikawa:2016znu,Monahan:2017hpu,Stewart:2017tvs} and the necessary matching of quasi-PDFs to the light-cone PDFs has been worked out~\cite{Ji:2017oey,Ji:2017rah}.

One of the issues quasi-PDFs calculations have to face is that the matrix elements have to be performed with large momentum hadron states. The momentum has to be large enough in order for the non-perturbative effects to be suppressed,   allowing for the  perturbative  matching formulas (currently one loop) to work. 
As discussed in~\cite{Radyushkin:2016hsy,Radyushkin:2017cyf}, non-perturbative evolution may dominate the approach of quasi-PDFs to the PDF limit.
The understanding that these non-perturbative effects are related to the transverse structure of the hadron led to a new suggestion of how one may extract PDFs from lattice QCD~\cite{Radyushkin:2017cyf}.  In this talk, results from our first numerical exploration~\cite{Orginos:2017kos} are presented.
%----------------------------------------------------------------------------

%  \begin{figure}[t]
%    $$
%  \includegraphics[width=0.5\textwidth]{disperkey.pdf} 
%  $$
%  \caption{Nucleon dispersion relation. The solid line is the continuum dispersion relation and the  error-band is an indication of the statistical error of the lattice nucleon energies. }
%  \label{fig:disp}
%\end{figure}
 
\section{Formalism }\label{sec-1}
In order to compute parton distribution functions in lattice QCD one may start from the equal time  hadronic matrix element with the quark and anti-quark fields separated by a finite distance. 
For  non-singlet parton densities   the matrix element
    \begin{align}
 {\cal M}^\alpha  (z,p) \equiv \langle  p |  \bar \psi (0) \,
 \gamma^\alpha \,  { \hat E} (0,z; A) \tau_3 \psi (z) | p \rangle \  , 
\label{Malpha}
\end{align}
is introduced.
 Here  ${ \hat E}(0,z; A)$ is  the   $0\to z$ straight-line gauge link 
 in the fundamental representation,  $\tau_3$ is the flavor Pauli matrix,  and $\gamma^a$ is an appropriate gamma matrix.
 By Lorentz invariance, this matrix element can  be decomposed as
\begin{align} 
{\cal M}^\alpha  (z,p) = &2 p^\alpha  {\cal M}_p (-(zp), -z^2) 
 + z^\alpha  {\cal M}_z (-(zp),-z^2)
\ .
\end{align}
From the ${\cal M}_p (-(zp), -z^2) $ part  the twist-2 contribution to PDFs  can be obtained in the limit  $z^2 \to 0$.
By taking    $z=(0, 0, 0, z_3)$,  $\alpha$ in the temporal direction i.e. \mbox{$\alpha=0$}, and the hadron momentum $p=(p^0,0,0,p)$  the  $z^\alpha$-part drops out.
In addition, we define  the Lorentz invariant $\nu = -(zp)$, which is known in the literature as the Ioffe time~\cite{Ioffe:1969kf,Braun:1994jq}.
With these definitions
 \begin{equation} 
 \langle  p |  \bar \psi (0) \,
 \gamma^0 \,  { \hat E} (0,z; A) \tau_3 \psi (z) | p \rangle \ 
 = 2 p^0 {\cal M}_p (\nu, z_3^2) \,.
 \label{eq:matelem}
\end{equation}
It should be noted here that  the  quasi-PDF $Q(y,p)$ is  related to ${\cal M}_p (\nu,z_3^2) $ by
\begin{align} 
  Q(y,  p)   =\frac{1}{2 \pi}  \int_{-\infty}^{\infty}  d\nu \, 
   \, e^{-i y  \nu}  \, {\cal M}_p (\nu,  [\nu/p]^2)    \  .
\label{IxM}
\end{align}  

Furthermore,  we introduce the Ioffe time PDFs $\mathcal{M}(\nu,z_3^2)$ defined at a scale $\mu^2 = 1/z_3^2$ which are the Fourier transform of regular PDFs $f(x,\mu^2)$ as
\begin{equation}
\mathcal{M}(\nu,z_3^2) =  \int_{-1}^1 dx \, f(x,1/z_3^2) e^{i x\nu} \, .
\end{equation}
The scale dependence of the Ioffe time PDF can be derived from the DGLAP evolution of the regular PDFs. The Ioffe time PDFs satisfy the following evolution equation
  \begin{align}
    \frac{d}{d \ln z_3^2} \,  
{\mathcal M} (\nu, z_3^2)    &= - \frac{\alpha_s}{2\pi} \, C_F
\int_0^1  du \,   B ( u )   {\mathcal M} (u \nu, z_3^2)\ , \;\; {\rm with}\;\; B (u)    =  \left [ \frac{1+u^2}{1- u} \right ]_+  \,, 
\label{EE}
 \end{align}
where  $C_F=4/3$, and $B(u)$ is  the leading-order evolution kernel  for the non-singlet quark PDF \cite{Braun:1994jq}.
  
 The $[ \ldots ]_+$  denotes the  ``plus'' prescription, i.e. 
    \begin{equation}
 \int_0^1  du \,    \left [ \frac{1+u^2}{1- u} \right ]_+     {\mathcal M} (u \nu)  
\nonumber\\  =
\int_0^1  du \,  \frac{1+u^2}{1- u} \,    [{\mathcal M} ( \nu)-{\mathcal  M} (u \nu)  ]\ . 
\label{plus}
 \end{equation}

       \begin{figure}[t]
       \vspace{-6mm}
 \includegraphics[width=0.45\textwidth]{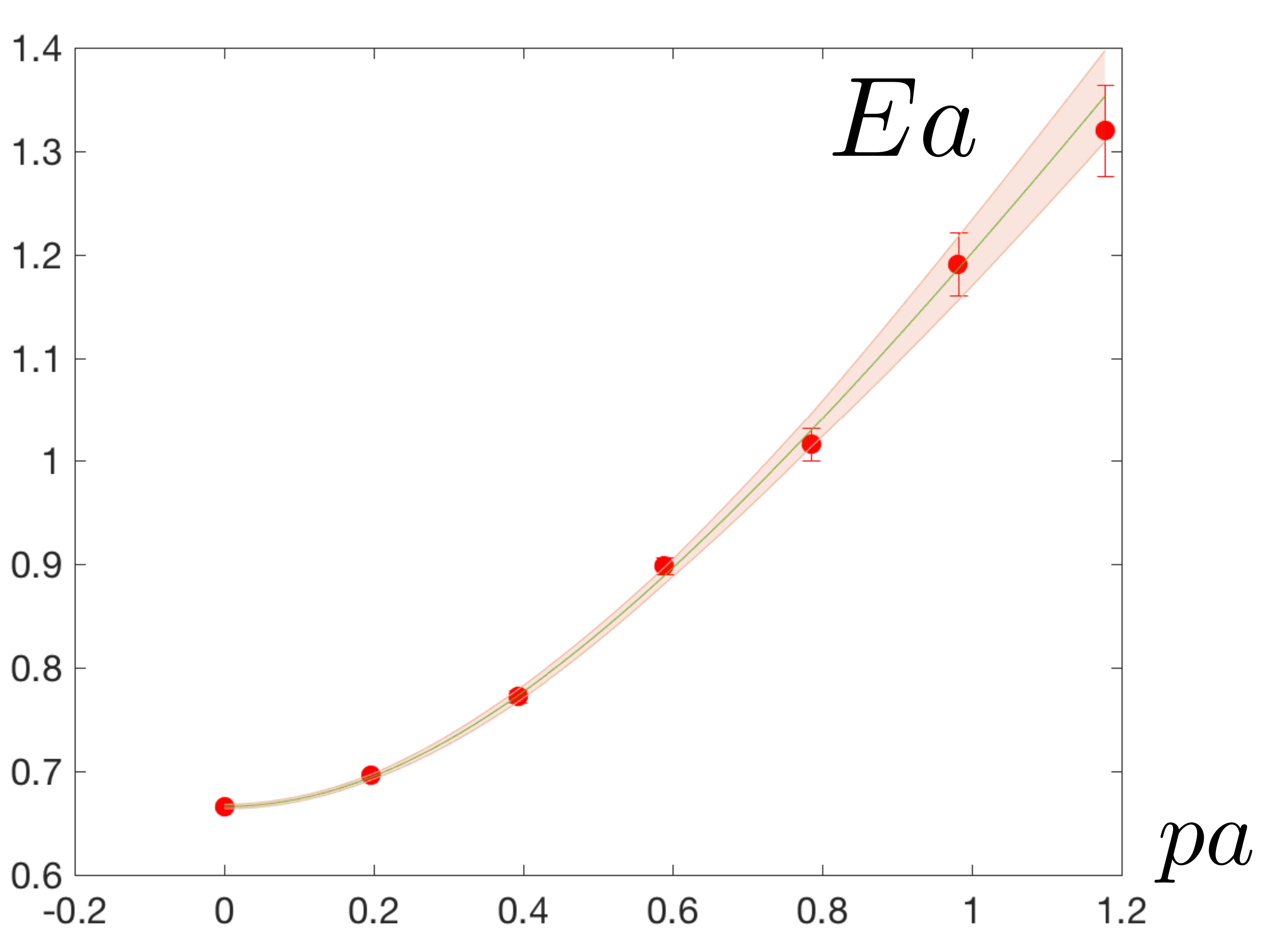} \hfill \includegraphics[width=0.45\textwidth]{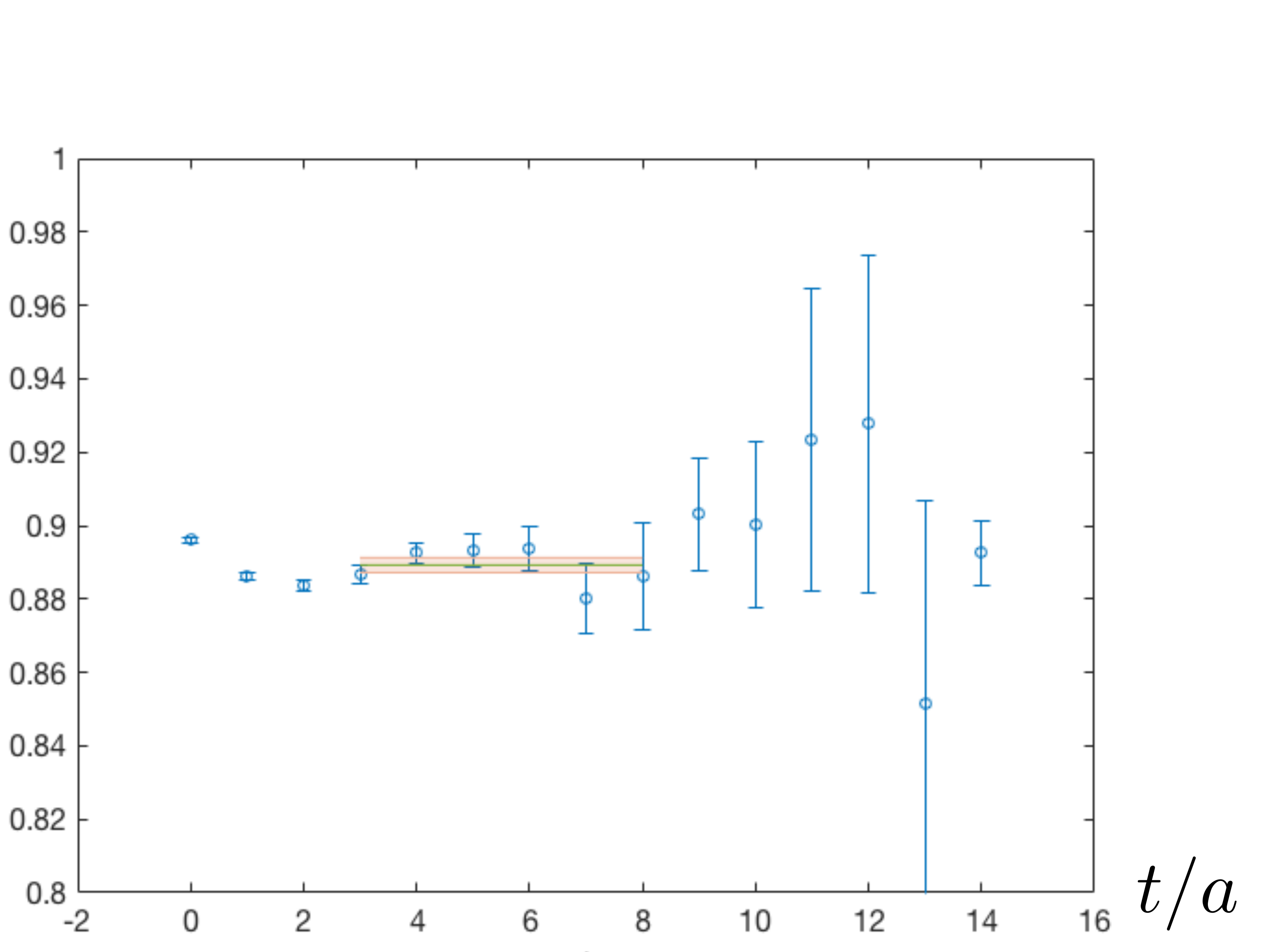}
  \caption{{\bf Left:} Nucleon dispersion relation.  {\bf Right:} Typical fit used to extract the reduced matrix element (here  $p=2\pi/L \cdot 2$ and $z=4$ ).
  }
  \label{fig:plateau}
\end{figure}

The Ioffe time PDFs can now be obtained from the matrix element of Eq.~\ref{eq:matelem} in the small $z_3$ limit as
 \begin{equation}
\mathcal{M}_p(\nu,z_3^2) =  \mathcal{M}(\nu,z_3^2) + \mathcal{O}(z_3^2)\,.
\label{eq:matelem_small_z}
\end{equation}
However, the corrections of $ \mathcal{O}(z_3^2)$ are large rendering the above relation practically useless.
On the other hand by the conservation of the vector current one knows
that
\begin{equation}
\mathcal{M}_p(0,z_3^2) =  1 + \mathcal{O}(z_3^2)\,.
\label{eq:matelem_small_z_nu0}
\end{equation}
As was argued in~\cite{Radyushkin:2017cyf} the $z_3^2$ corrections in both Eq.~\ref{eq:matelem_small_z} and Eq.~\ref{eq:matelem_small_z_nu0}
are related to the transverse structure of the hadron and hopefully cancel approximately in a ratio. Therefore it was argued that
\begin{equation}
{\mathfrak M} (\nu, z_3^2) \equiv \frac{ {\cal M}_p (\nu, z_3^2)}{{\cal M}_p (0, z_3^2)} %= \mathcal{M}(\nu,z_3^2) + \mathcal{O}(z_3^2) 
 \,,
\label{eq:ratio_small_z}
\end{equation}
has much smaller  $ \mathcal{O}(z_3^2)$ corrections  and therefore this ratio could be used to extract the Ioffe time PDFs in a more practical manner.
Furthermore the ratio in Eq.~\ref{eq:ratio_small_z} has a well defined continuum limit and does not require renormalization.
Therefore, it is well suited for lattice QCD calculations.

\section{Numerical tests}\label{sec-2}

  We performed lattice   QCD calculations in the quenched approximation at $\beta=6.0$ on $32^3\times 64 $
  lattices, with the non-perturbatively tuned clover fermion action with the
   clover coefficients computed by the Alpha  \mbox{collaboration \cite{Luscher:1996ug}.}
   Correlation functions were computed on a total of 500 configurations separated by 1000 updates, each one consisting of four over-relaxation   and one 
   heatbath sweeps.  On each configuration we computed correlation functions from six randomly selected smeared around a  point sources.
   The pion  and nucleon masses  determined to be $601(1)$ MeV  and $1411(4)$MeV respectively. The lattice spacing in our calculation is  $a=0.093$ fm 
   as determined by the Alpha collaboration~\cite{Necco:2001xg}. 
   
   The maximum  nucleon momentum used in our calculation was $2.5\, $GeV.
   Inside  this momentum  range, the continuum dispersion relation for the nucleon was satisfied within the errors of the calculation,
    indicating small lattice artifacts of ${\cal O}(ap)$.
    In figure \ref{fig:plateau}(left) we plot the nucleon energy as a function of   momentum along with the continuum dispersion relation.

  The computation of the matrix elements was performed using 
   the methodology described in~\cite{Bouchard:2016heu} with an operator insertion given by Eq.~(\ref{Malpha}). 
   
  Following~\cite{Bouchard:2016heu} we need to compute  a regular nucleon two point function given by
  \begin{equation}
  C_{p}(t) = \langle \mathcal{N}_{p}(t)\overline{ \mathcal{N}}_{p}(0) \rangle\,  ,
  \end{equation}
where $ \mathcal{N}_{p}(t)$ is a helicity averaged, non-relativistic nucleon interpolating field with momentum $p$. The quark fields in $ \mathcal{N}_{p}(t)$ are smeared with a gauge  invariant Gaussian smearing. This choice of an interpolation field 
is known to couple well to the nucleon ground state (see discussion in~\cite{Bouchard:2016heu}).
 The quark smearing width was optimized to give good  {overlap with the nucleon ground state within the}  range of momenta {in our calculation}.  In addition we need the correlator given by
  \begin{equation}
  C^{\mathcal{O}^0(z)}_p(t) =\sum_\tau \langle \mathcal{N}_p(t) \mathcal{O}^0(z,\tau)\overline{ \mathcal{N}}_p(0) \rangle\; \; \; {\rm with}\;\;\; \mathcal{O}^0(z,t)= \overline\psi(0,t) \gamma^0 \tau_3 \hat{E}(0,z;A)\psi(z,t)\, ,
  \end{equation}
 and $\tau_3$  being the flavor Pauli matrix.
  The proton momentum and the displacement of the quark fields were both taken along the $\hat z$ axis ($\vec z = z_3 \hat z$ and $\vec p = p \hat z$).
  We  define the effective matrix element as 
  \begin{equation}
  \mathcal{M}_{\rm eff}(z_3 p, z_3^2;t) = \frac{C^{\mathcal{O}^0(z)}_p(t+1)}{C_p(t+1)} - \frac{C^{\mathcal{O}^0(z)}_p(t)}{C_p(t) }  \  . 
  \end{equation}
  As it was shown in~\cite{Bouchard:2016heu}, the desired matrix element  $\mathcal{J}$  of Eq.~(\ref{Malpha}) can  be extracted at the large Euclidean time separation as
  \begin{equation}
\frac{\mathcal{J}(z_3p, z_3^2)}{2 p^0}= \lim_{t\rightarrow\infty}  \mathcal{M}_{\rm eff}(z_3 p, z_3^2;t) \  ,
  \end{equation} 
  where $p^0$ is the energy of the nucleon. 
 
The  resulting effective matrix element has contamination from excited states that scale as $e^{-t \Delta E }$,
where $t$ is the Euclidean time separation of the nucleon creation and annihilation operators,
and $\Delta E$ is the mass gap to the first excited state of the nucleon. This contamination can be fitted away and this is what we did in our calculations. 
This methodology  allows for the computation of  all nucleon matrix elements that correspond to different nucleon momentum  and spin polarization as well as different diagonal  flavor structures in the matrix element without additional computational cost. This fact, together with our emphasis on  having as many nucleon momentum states as possible,  makes the total computational cost of this approach less than the equivalent cost of performing the calculations with the sequential source method.   This approach has recently been successfully used for both single and multi-nucleon matrix element calculations~\cite{Berkowitz:2017gql,Tiburzi:2017iux,Shanahan:2017bgi}.

  % \newpage 
 %
In order to renormalize our lattice matrix elements we note that, for $z_3=0$, the matrix element 
 $\mathcal{M}(z_3 p, z_3^2)$  corresponds to a  local vector (iso-vector) current, and therefore should  be equal to 1. However, on the lattice this is not the case due to lattice artifacts.
 Therefore we introduce a renormalization constant 
  \begin{equation}
  Z_p = \frac{1}{ \left.\mathcal{J}(z_3 p, z_3^2)\right|_{z_3=0}}\, .
  \label{eq:renorm}
  \end{equation}
 The factor  $Z_p$ has to be independent from $p$.  However,  again due to lattice
   artifacts or potential fitting systematics,  this is not the case. For this reason,  
   we renormalize  the matrix element for 
   each momentum  with its own $Z_p$ factor taking this way advantage of maximal statistical correlations {to reduce statistical errors},  as well as the cancellation of lattice artifacts in the ratio.
  Therefore,  our  matrix element is extracted  using the double  ratio
  \begin{equation}
\mathfrak{M}(\nu, z_3^2)=  \lim_{t\rightarrow\infty} 
 \frac{\mathcal{M}_{\rm eff}(z_3 p, z_3^2;t)}{\left.\mathcal{M}_{\rm eff}(z_3 p, z_3^2;t)\right|_{z_3=0}} 
 \times 
\frac{\left.\mathcal{M}_{\rm eff}(z_3  p, z_3^2;t)\right|_{p=0,z_3=0}}
{\left.\mathcal{M}_{\rm eff}(z_3  p, z_3^2;t)\right|_{p=0}}  \ ,
\label{eq:redMatElem}
  \end{equation} 
  which takes care of the renormalization of the vector current according to Eq.~(\ref{eq:renorm}).
  In practice, the   infinite $t$ limit is obtained with a fit to a constant  for  a suitable choice of a fitting range. In all cases we studied,  the average $\chi^2$ per degree of freedom was ${\cal O}(1)$. Typical fits used to extract the reduced matrix element are presented in Fig.~\ref{fig:plateau}. All fits are performed with the full covariance matrix and the error bars are determined with  the jackknife method. The reduced  matrix element defined in Eq.~(\ref{eq:redMatElem}) has a well defined continuum limit and no additional renormalization is required. This continuum limit is obtained at fixed $\nu$ and $z^2$ as well as at fixed quark mass. 

\subsection{Results}

    \begin{figure}[t]
    \centerline{\includegraphics[width=.48\textwidth]{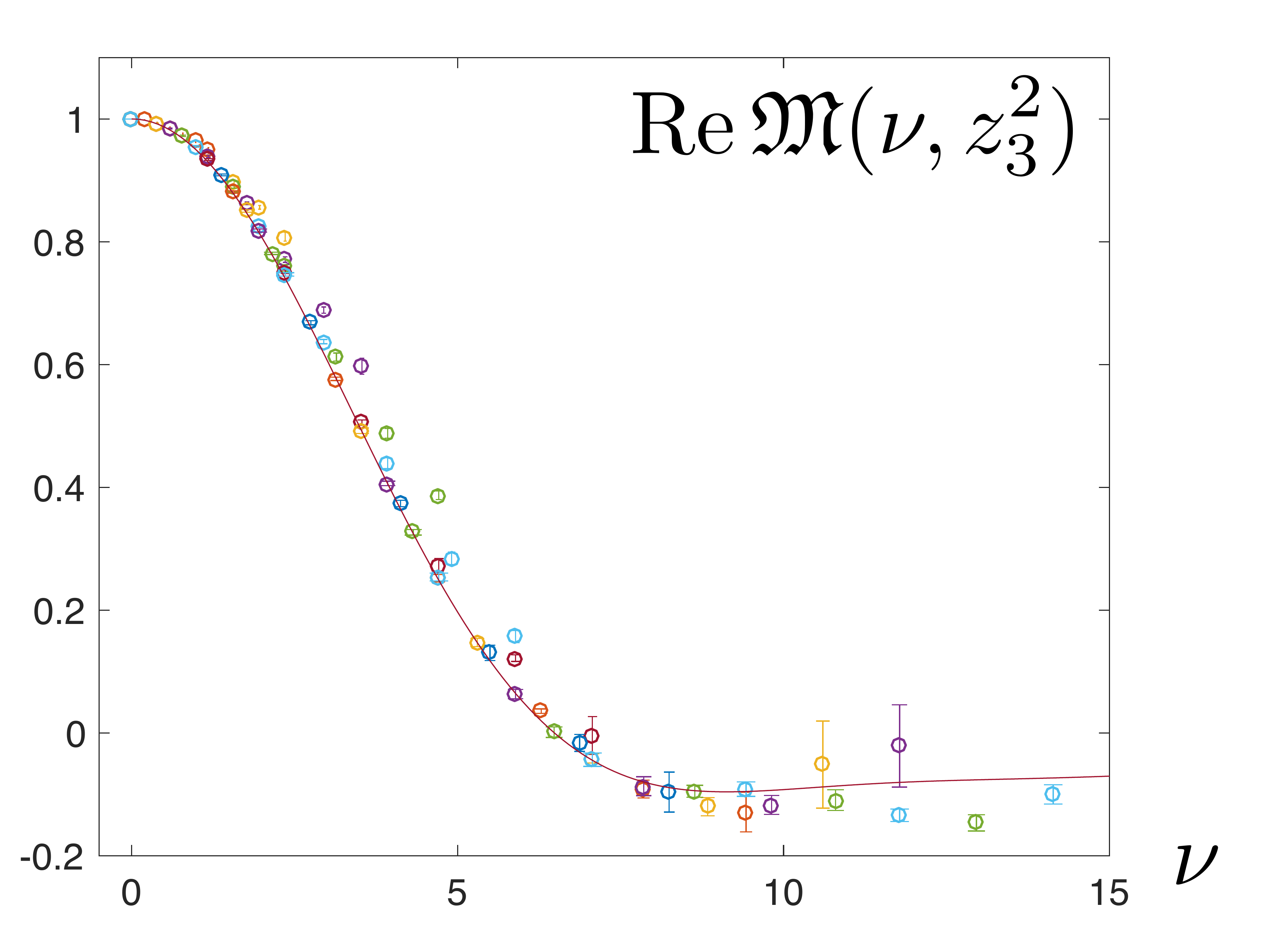} \hfill {\includegraphics[width=.48\textwidth]{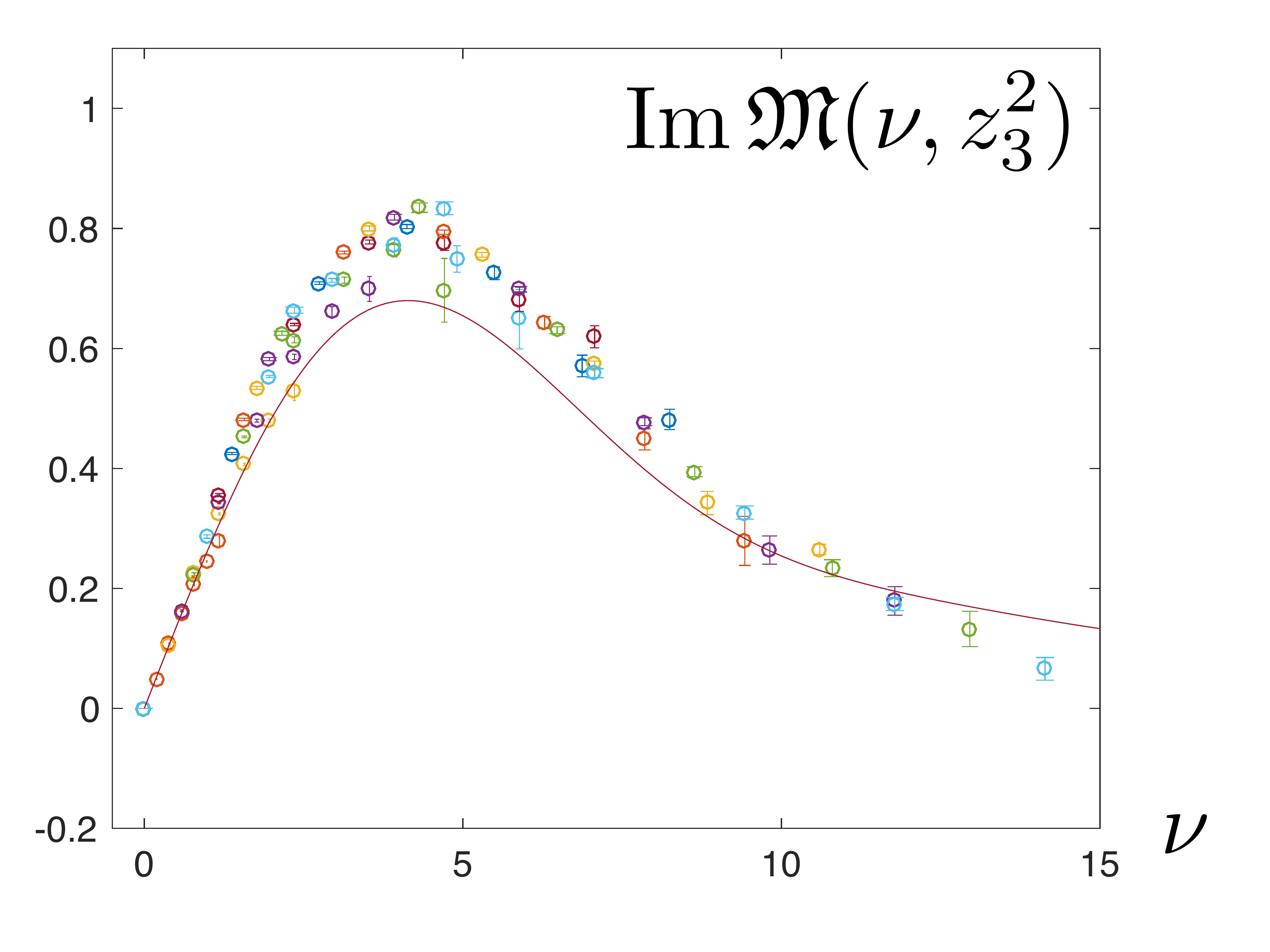} }}
   % \vspace{-0.6cm}
    \caption{ Real  and Imaginary parts of ${\mathfrak  M} (\nu, z_3^2)$. The  curves are for comparison and are  given by Eqs. (\relax {\ref{MC}}), (\relax {\ref{qV}}).
        \label{realc}}
    \end{figure}

In  Fig. \ref{realc} , we plot  the  ratio $ \mathfrak{M}(\nu, z_3^2)$  as a function of the Ioffe time $\nu$.
As one can see,  the  data approximately fall on the same curve.
For  the  imaginary part, the situation is similar.  This phenomenon can be understood if an approximate  factorization of the 
longitudinal and transverse structure of the hadron occurs.

    The real part  ${\mathfrak  M}_R (\nu)$ of the Ioffe-time distribution is obtained from  the cosine Fourier transform 
  \begin{align} 
{\mathfrak  M}_R (\nu) \equiv   & \int_0^1 dx \,  
\cos (\nu x)  \, q_v  (x)  
\label{MC}
     \end{align}
of the function  $q_v(x)$  given by the 
difference  $q_v(x) = q(x) -\bar q(x)$  of 
quark and antiquark   distributions. 
 In our case,   $q$ is  $u-d$ and 
    $\bar q = \bar u - \bar d$ and thus the $x$-integral of $q - \bar q$     in the proton should be equal to 1. 

The  imaginary part   ${\mathfrak  M}_I (\nu)$   of the Ioffe-time distribution is given by the sine Fourier transform 
      \begin{align} 
  {\mathfrak  M}_I (\nu)  \equiv
& \int_0^1 dx \, 
\sin (\nu x) \, q_+  (x) 
     \end{align} 
   where  the function $q_+ (x)= q(x) + \bar q (x)$,
   which may be also represented as \mbox{$q_+ (x)= q_v(x) + 2 \bar q (x)$ }.
   In order to guide the eye  we also plotted the corresponding Ioffe time distribution resulting from a PDF with the following form
     \begin{align} 
    q_v(x) =  \frac{315}{32} \sqrt{x}  (1-x)^{3}\,. 
    \label{qV}
    \end{align}
   If we neglect the antiquark contribution and use \mbox{$q_+  (x) = q_v(x)$,} then both the real and imaginary parts are obtained by the cosine and sine Fourier transforms of the above function. This function was chosen to result in a curve that closely follows the data for the  real part. 
   As argued in~\cite{Orginos:2017kos} the fact that the imaginary part data seem not  to be  following  the  sine transform curve is an indication of non-zero anti-quark densities.
 
   By looking more closely at the data plotted as a function of the Ioffe time we can see that 
  there is  a residual \mbox{$z_3$-dependence}. This is more visible when, for a particular $\nu$,    there  are several  data points 
  corresponding to different values of $z_3$. This is expected because as discussed above different values of $z_3^2$ for the same $\nu$ correspond to the Ioffe time distribution at different scales. It is interesting to check whether the residual scatter in the data points is consistent with evolution.
By solving the evolution equation at leading order,  the Ioffe time PDF at $z'_3$ can be related to that at $z_3$ by the following relation
  \begin{align}
{\mathfrak M} (\nu, {z'}_3^2) {=}
{\mathfrak M} (\nu, z_3^2) \
 -\frac23  \frac{\alpha_s}{\pi}   \ln ({z'_3}^2/z_3^2) \int_0^1 du\, B(u) \, {\mathfrak M}\,  (u\nu, z_3^2)\, .
 \label{Prop}
 \end{align}
  This formula is only applicable at small $z_3$ and therefore we check its effect to our data at values of $z_3\le4 a$ which correspond to energy scales larger than 500 MeV.
 More specifically, we fix  the point  $z'_3$ at the   value \mbox{$z_0=2a$}   corresponding, 
 at the leading logarithm level,  %%%%%%%%
 to the
 $\overline {\rm MS}$-scheme scale \mbox{$\mu_0 = 1$  GeV }  and evolve the rest of the points to that scale.

 % \begin{figure}[t] 
   % \vspace{-0.6cm}
%    \caption{Real    part of ${\mathfrak  M} (\nu, z_3^2)$ 
%    for $z_3/a =1, 2, 3,$ and 4. {\bf Left:} Data before evolution. {\bf Right:} Data after evolution. The reduction in scatter indicates that evolution indeed collapses all data to the same universal curve.
%        \label{ER}}
 %   \end{figure}

\begin{figure}[t]
    \centerline{ \includegraphics[width=0.46\textwidth]{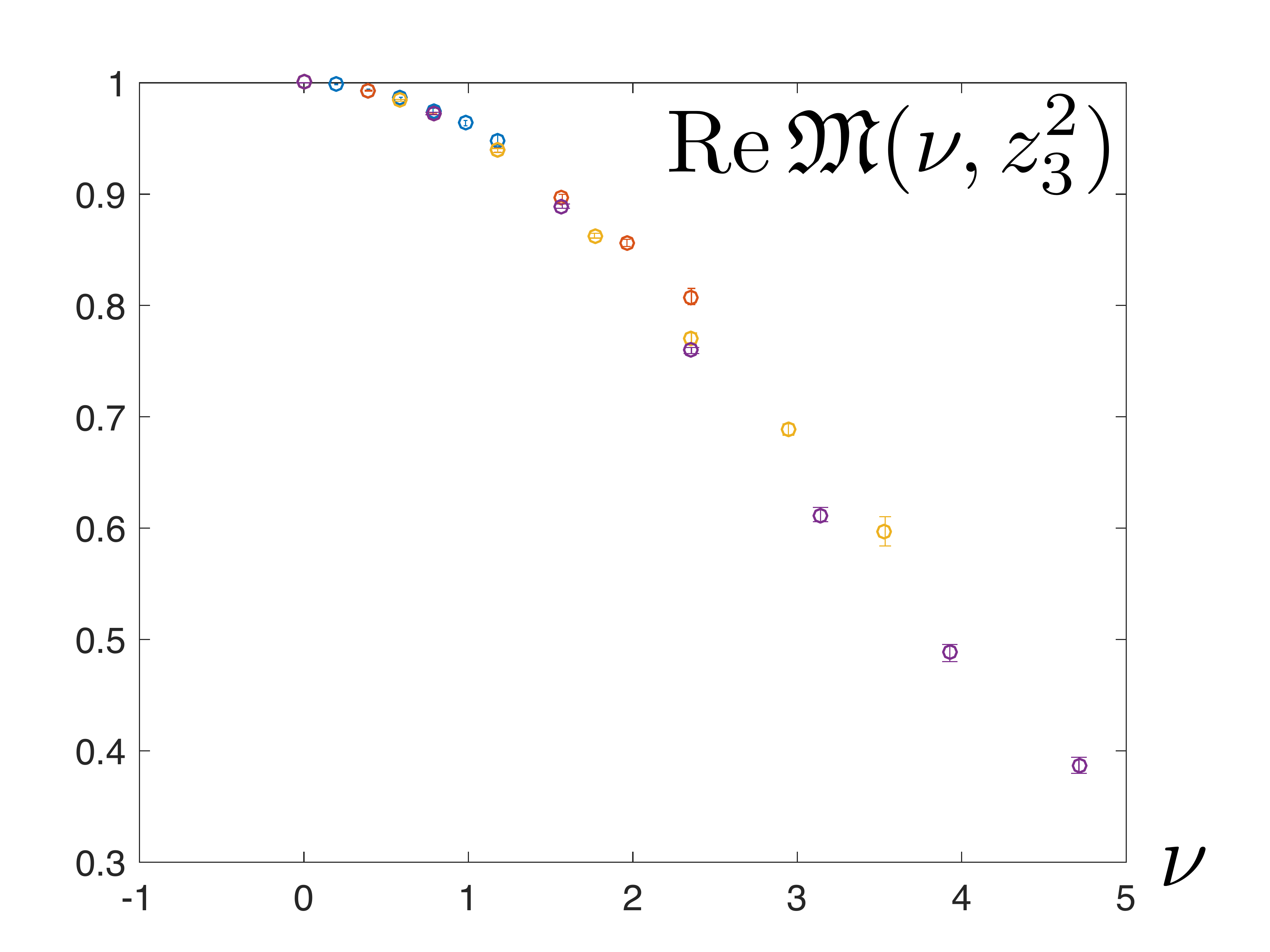}\hfill \includegraphics[width=0.46\textwidth]{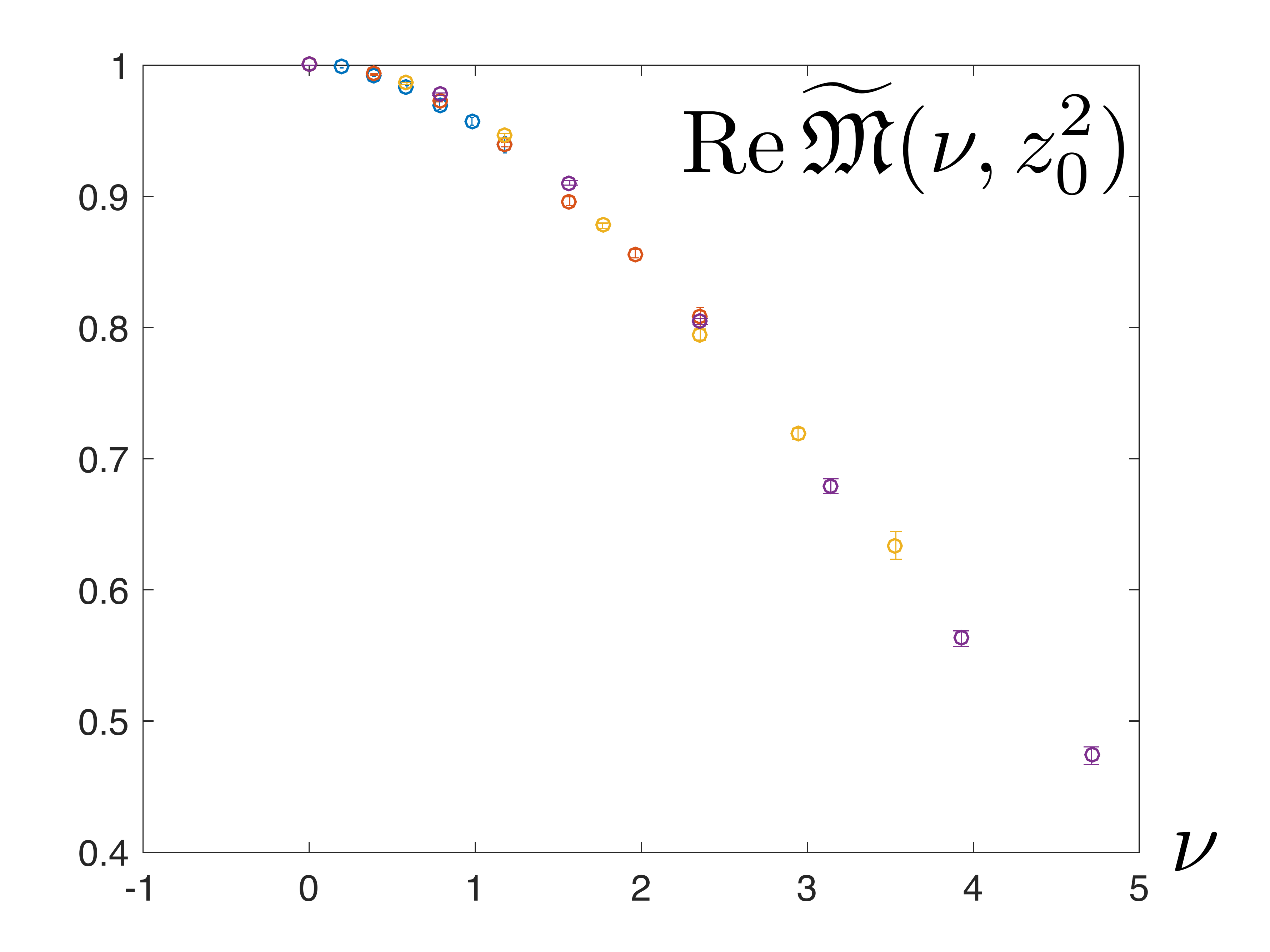}}
    \centerline{ \includegraphics[width=0.46\textwidth]{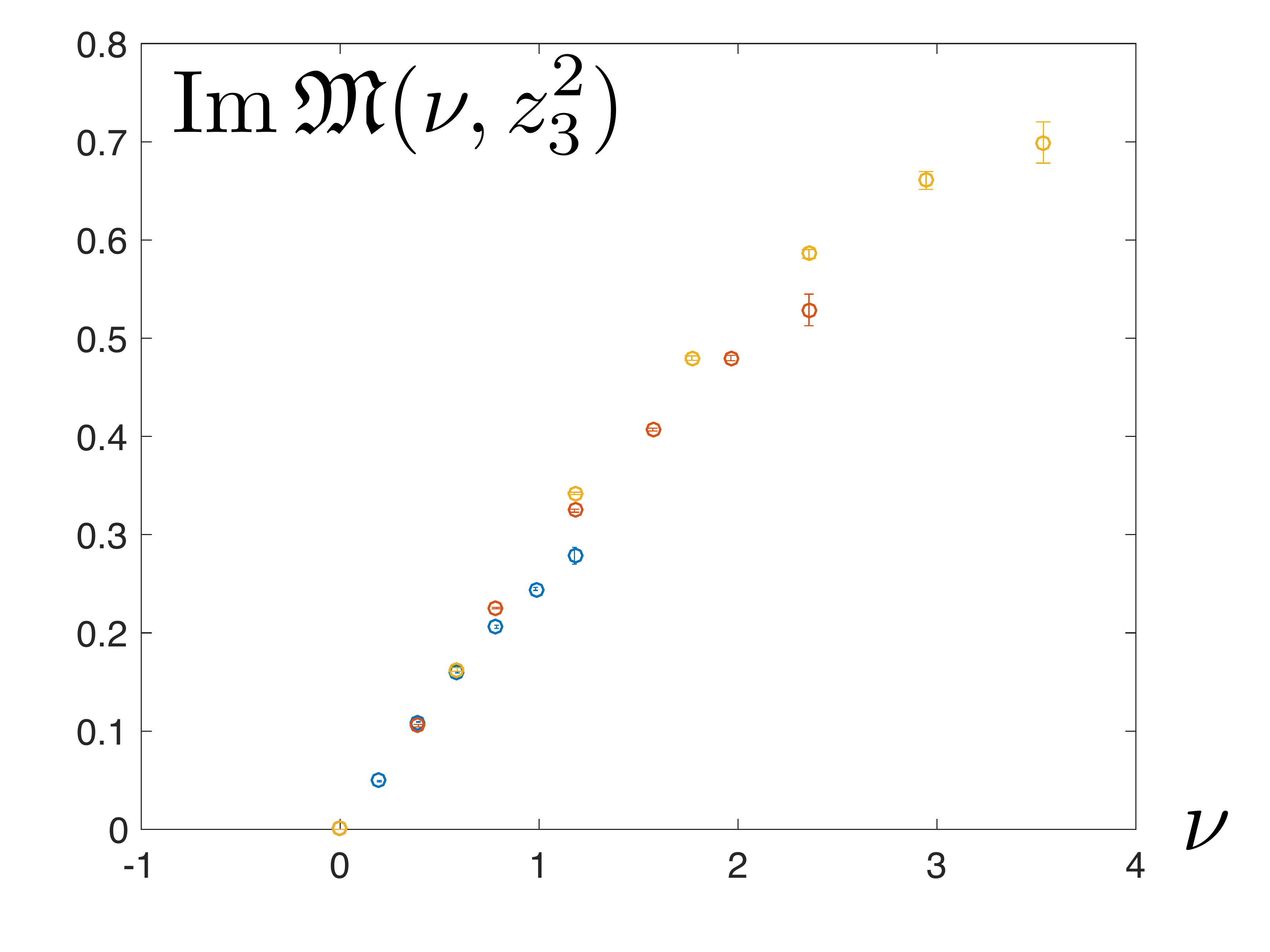}\hfill \includegraphics[width=0.46\textwidth]{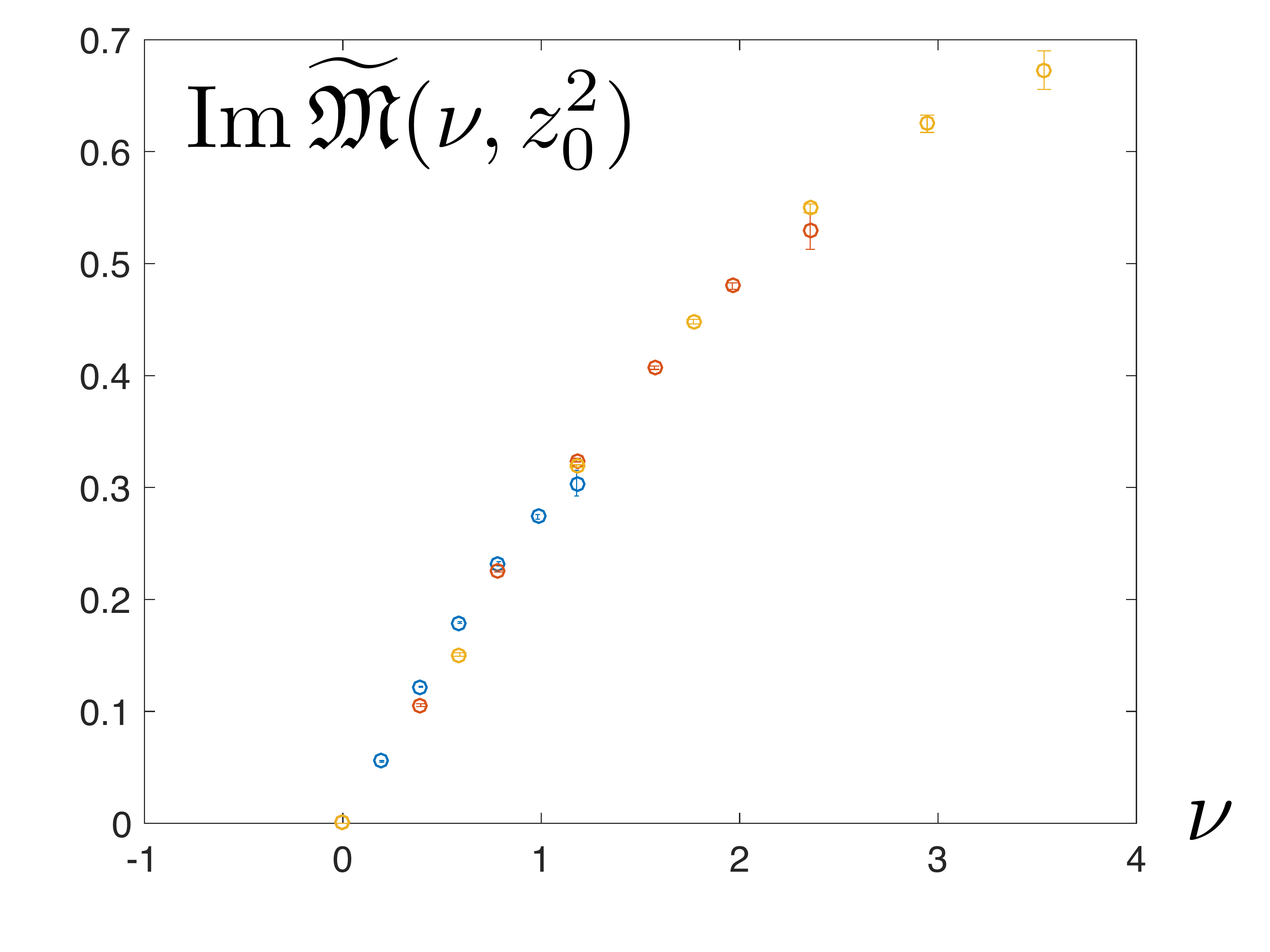}}
   % \vspace{-0.6cm}
    \caption{ The ratio  ${\mathfrak  M} (\nu, z_3^2)$  for 
    for $z_3/a =1, 2, 3,$ and 4. {\bf: Top:} Real part. {\bf Bottom: } Imaginary part.  {\bf Left:} Data before evolution. {\bf Right:} Data after evolution. The reduction in scatter indicates that evolution collapses all data to the same universal curve.
        \label{ER_EI}}
    \end{figure}
  % \newpage 
 
      \begin{figure}[t]
    \centerline{ \includegraphics[width=0.46\textwidth]{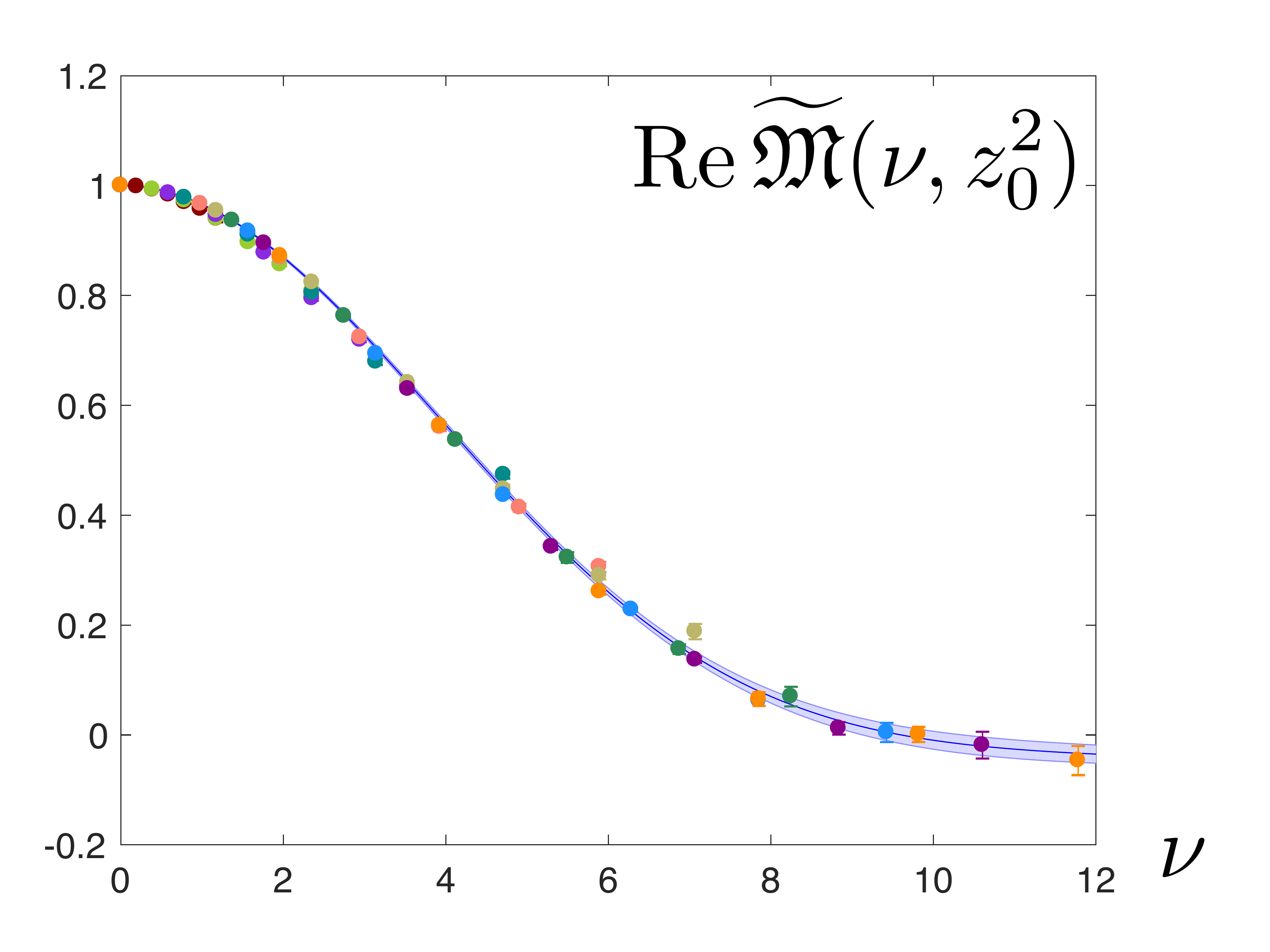}\hfill \includegraphics[width=0.46\textwidth]{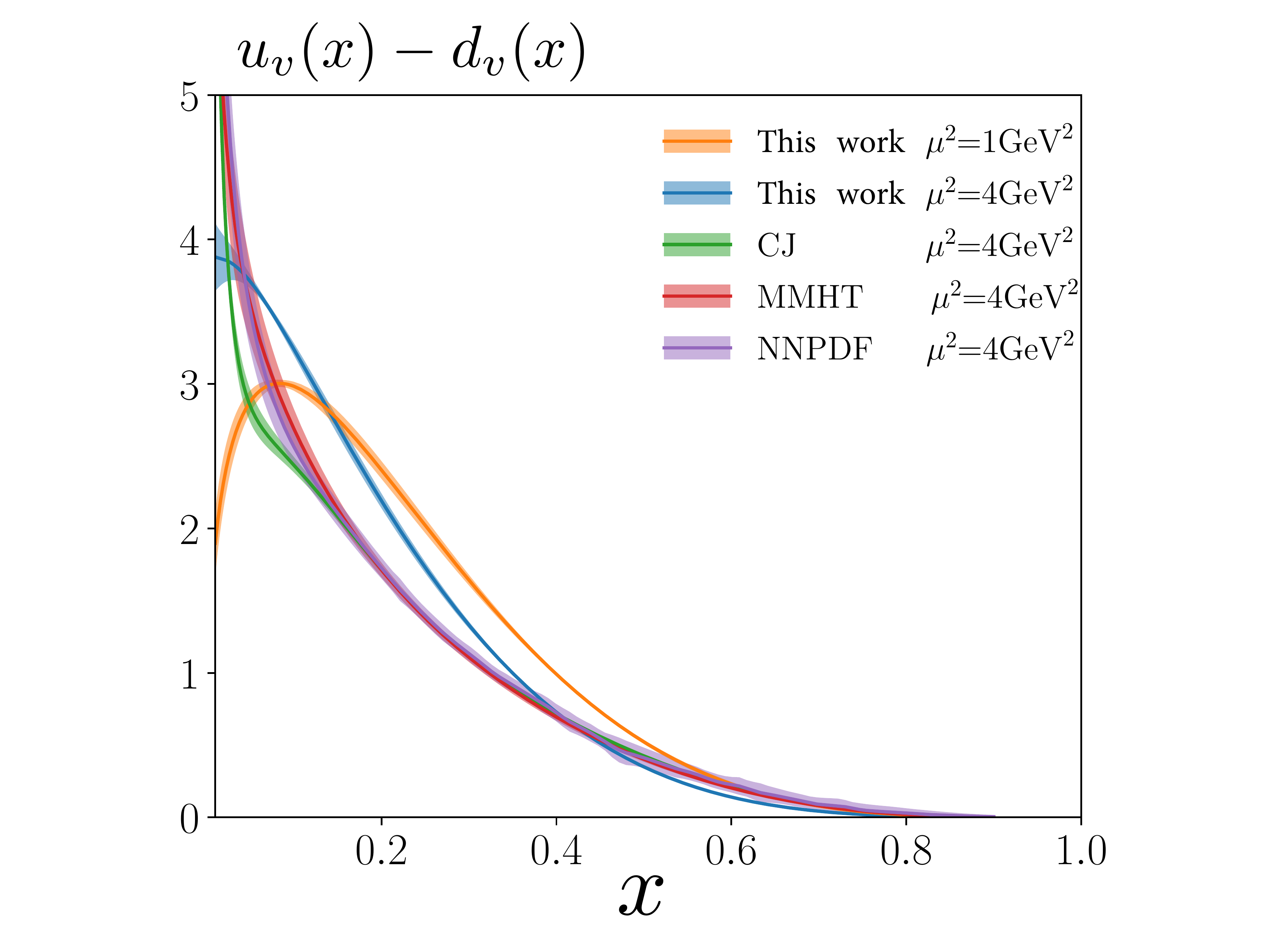} }
    \vspace{-0.3cm}
    \caption{Left: Data points for Re ${\mathfrak M}\,  (\nu, z_3^2)$ with $z_3 \leq 10a$ 
     evolved to $z_3=2a$ as described in the text. Right: Curve for $u_v(x) - d_v(x)$ built from the evolved data
   shown in the left panel and     treated as corresponding to
     the  \mbox{$\mu^2 =1$  GeV$^2$}   scale; 
then  evolved to the reference point  $\mu^2 =4$  GeV$^2$  of the  global fits.
        \label{evpoi}}
    \end{figure}

The real part data of our Ioffe time PDF are shown in  left panel of  \mbox{Fig. \ref{ER_EI} (top). }
 As one can see, there is a visible scatter of the data points.  
    Using $\alpha_s/\pi =0.1$, we calculate the
    ``evolved'' data points corresponding to  the function $\widetilde{\mathfrak M} (\nu, z_0^2) $.
    The results are shown in  the right panel of Fig. \ref{ER_EI} (top).  The evolved data points  are now very close
    to a universal  curve indicating that the original scatter was due to the evolution of the Ioffe time PDF with $z_3^2$.
The imaginary  part data of our Ioffe time PDF,  which are shown in \mbox{Fig. \ref{ER_EI} (botom)} (left: original data, right: evolved),  were evolved in the same manner, also support this claim.

It is reasonable to expect that leading order evolution cannot be extended to very low scales. However, it is known that evolution stops below a certain scale. By observing our data we can deduce that this is indeed the case for  length scales $z_3 \geq 6a$. We therefore adopt an evolution that leaves the PDF unchanged for length scales above  $z_3=6a$ and use the leading perturbative evolution formula to evolve to smaller $z_3$ scales. The resulting evolved data are presented in Fig.~\ref{evpoi}.
  By fitting  these evolved  points with a cosine Fourier transform ${\cal M} (\nu;a,b)$ of the normalized  valence PDF $q_v(x)$ satisfying 
%\begin{equation}
$q_v(x) = N(a,b) x^a(1-x)^b$
%\label{eq:fit_qv}
%\end{equation} 
we obtain $a = 0.36(6)$  and $b=3.95(22)$, where the errors are statistical only. This fit with its corresponding error-band is also plotted in Fig.~\ref{evpoi}.

Treating $z_3=2a$ as the $\overline{\rm MS}$ scale $\mu=1$ GeV, one can further evolve  the curve to the standard reference scale $\mu^2 = 4$ GeV$^2$ of the global fits, see right panel of Fig. \ref{evpoi}  
\footnote{We are very grateful 
to Nobuo Sato who performed this numerical evolution and provided the figure.}.  %%%%%%%%%%%
Our resulting curve follows closely the phenomenologically expected behavior. Given our limited range of $\nu$ the small x results are prone to larger systematic errors that need to be studied more carefully.  However, it is clear that in order to obtain results that reproduce the experimentally determined PDFs
one needs to perform more realistic dynamical fermion calculations including  quarks  with physical masses as well as treat evolution at higher accuracy than the one we used here.

%----------------------------------------------------------------------------
\section{Summary}\label{sec:details}

In this talk a new approach for obtaining PDFs from lattice QCD calculations is presented. We introduce  a ratio of matrix elements that takes care of UV divergences allowing 
for a well defined continuum limit. In addition, this ratio has improved convergence properties to the light-cone limit allowing for a practical method for performing these calculations with realistic computational resources. We tested this approach in the quenched approximation in order to  understand the basic features of the method and work out  the details of the methodology. An important finding of our calculations is that our data are consistent with the well known scale evolution of the PDFs. Armed with the lessons obtained by the quenched approximation, we are currently applying this approach to realistic lattice QCD calculations aiming towards obtaining a  precise determination of PDFs from lattice QCD.
 
 \section*{Acknowledgments}

 One of us (AR) thanks V. Braun and X. Ji for discussions and comments. 
We are indebted to Nobuo Sato for the help in comparison of our results with global fits.
This work is supported by Jefferson Science Associates,
 LLC under  U.S. DOE Contract \#DE-AC05-06OR23177. 
 KO and JK  was supported in part by U.S.  DOE grant
\mbox{ \#DE-FG02-04ER41302}, and in part by STFC consolidated grant ST/P000681/1. 
 AR was   supported in part 
 by U.S. DOE Grant   \mbox{\#DE-FG02-97ER41028. }
 SZ acknowledges support
by the National Science Foundation (USA) under grant
PHY-1516509. 
JK was supported in part by a DOE SCGSR fellowship at JLab.
 This work was performed in part using computing facilities at the
  College of William and Mary which were provided by contributions from the 
  National Science Foundation (MRI grant PHY-1626177), the Commonwealth of 
  Virginia Equipment Trust Fund and the Office of Naval Research. In addition,  
  this work used resources at NERSC, a DOE Office of Science User Facility supported by the Office of Science of the U.S. Department of Energy under Contract \#DE-AC02-05CH11231.

 \clearpage
\bibliography{lattice2017}

%%%%%%%%%%%%%%%%%%%%%%%%%%%%%%%%%%%%%%%%%%%%%%%%%%%%%%%%%%%%%%%%%%%%%%%%%%%%%
\end{document}